\documentclass{article}  
\usepackage{spadre2008}
\usepackage{graphicx}
\usepackage{xspace}
\frompage{000} \topage{000}                                              
\def\rightmark{PHENIX Jets vs.\ Reaction Plane}
\def\leftmark{D. Winter}
 
\title{PHENIX Results on Jet Correlations versus Reaction Plane} 
\authors{
{David L. Winter$^1$, for the PHENIX Collaboration %
}\\[2.812mm]
{\normalsize
\hspace*{-8pt}$^1$ Columbia University, \\ 
New York, NY USA}}
 
\abstract{In relativistic heavy-ion collisions, studying jet
correlations and their dependence on the angle of emission with
respect to the reaction plane can be used to shed light on the path
length dependence of the energy lost by the jet.  In this paper, we
present recent PHENIX results on jet correlations versus reaction
plane, as a function of centrality and transverse momentum.}


\newcommand{\pT}     {\ensuremath{p_\mathrm{T}}\xspace}

\newcommand{\vTwo}   {\ensuremath{v_2}\xspace}
\newcommand{\dphi}   {\ensuremath{\Delta\phi}\xspace}
\newcommand{\dphirp} {\ensuremath{\Delta\phi_{RP}}\xspace}
\newcommand{\GeVc}   {\ensuremath{\mathrm{GeV}\!/c}\xspace}
\newcommand{\snn}    {\ensuremath{\sqrt{s_{NN}}}\xspace}
\newcommand{\pp}     {p+p\xspace}
\newcommand{\AuAu}   {Au+Au\xspace}
\newcommand{\CuCu}   {Cu+Cu\xspace}

\begin{document}
 
\maketitle
\setcounter{page}{1}

\section{Introduction}\label{intro}

Jet production from the hard-scattering of partons in relativistic
heavy ion collisions at RHIC is an important observable for the
strongly interacting matter created.  The observation of jet quenching
in particular has provided key insight into the produced medium and
its response to energy lost by fast partons as they traverse it.  The
partons' energy loss is predicted to be dependent on gluon densities
and the path length traversed by the jet.  Because of the difficulty
of full jet reconstruction in the high multiplicity environment of a
heavy ion collision, azimuthal correlations are a crucial tool for
extracting such jet properties.

\begin{figure}
\begin{center}
\includegraphics[height=4.4cm]{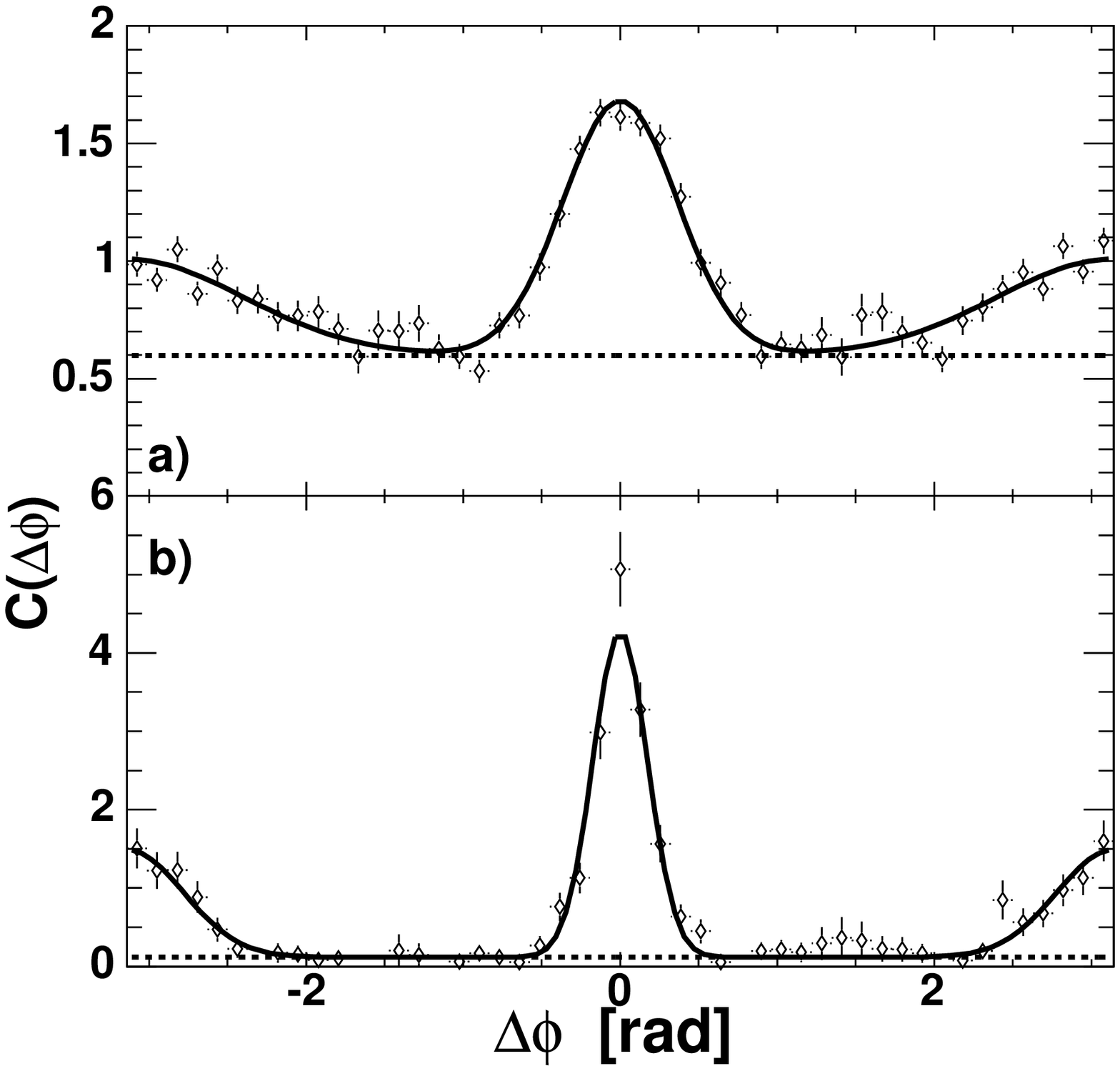}
\includegraphics[height=4cm]{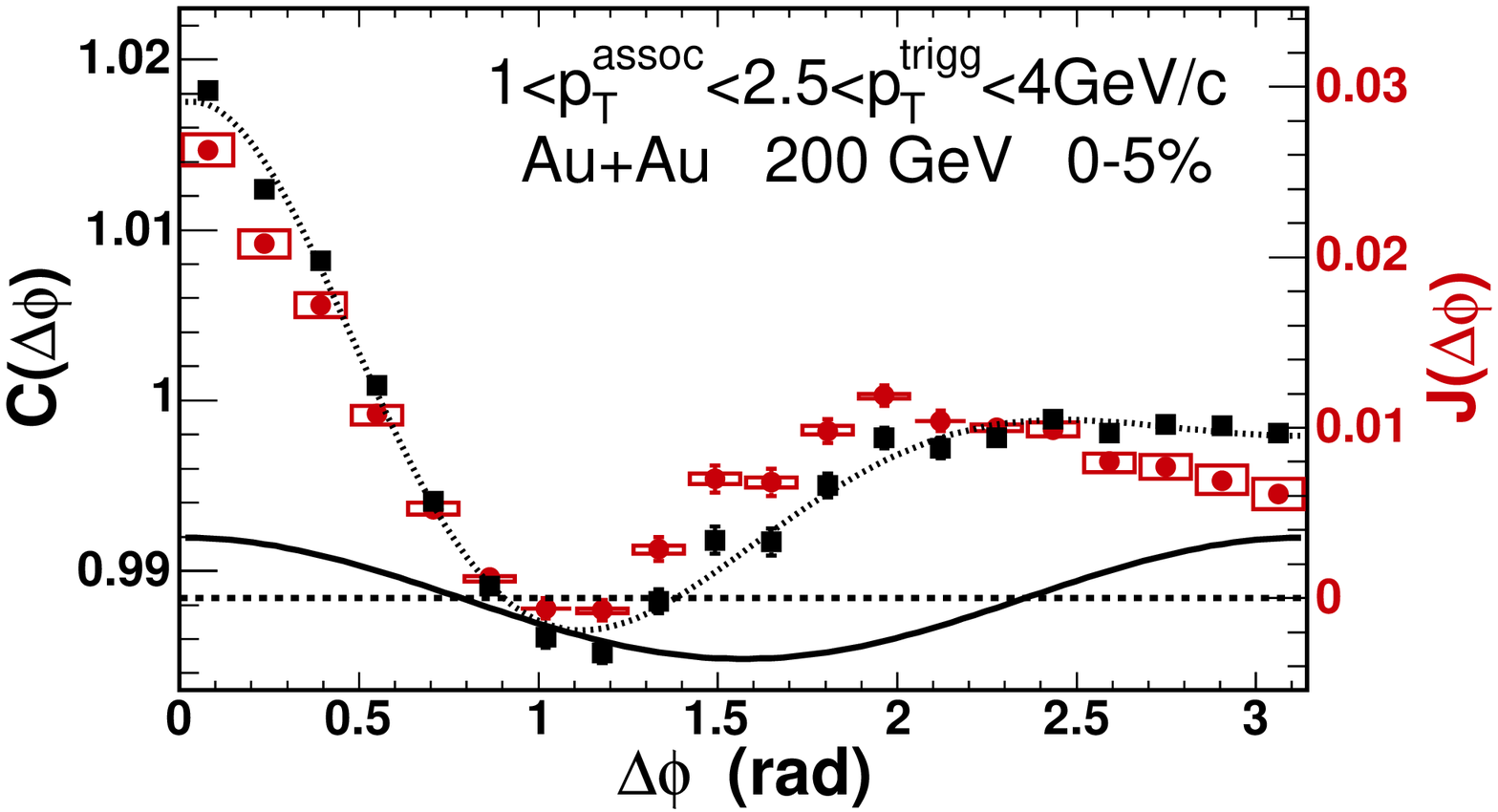}
\end{center}
\vspace*{-14pt}
\caption{Left: Measured two-particle correlations for \pp collisions
  at RHIC, for both particles in the momenta ranges of a)
  $1.5<\pT<2.0~\GeVc$ and b) $3.0<\pT<4.0~\GeVc$. Data and fits are
  described in Ref.~\cite{Adler:2006sc}.  Right: Measured correlation
  and jet functions for central \AuAu collisions at
  $\snn=200$~GeV~\cite{Adare:2006nr}.}
\vspace*{-14pt}
\label{fig:fig-1}
\end{figure}

For each event, correlations in azimuthal angle $\phi$ are constructed
by selecting the leading (high-\pT) particle (called the
\emph{trigger} particle) and other particles in the event (called the
\emph{associated} particles)\cite{:2008cq}.  Subsequent to
hard-scattering, the partons fragment into a cone of hadrons which are
the particles observed in the laboratory.  The trigger particle serves
as a proxy for the jet axis.  As a result, we expect to find peaks at
$\dphi = 0$ and $\dphi=\pi$ (where $\dphi = \phi_{trig} -
\phi_{assoc}$), corresponding to the jet of the trigger particle and
the momentum-balanced jet on the away-side respectively.  This is
illustrated in Figure~\ref{fig:fig-1}, and has been observed in \pp
collisions~\cite{Adler:2006sc}.
Also shown in Figure~\ref{fig:fig-1} is the two-particle correlation
function in heavy-ion collisions. 

A striking feature of the jet-induced angular distributions is the
modification of the away-side peak~\cite{:2008cq,Adler:2005ee} as seen
in Figure~\ref{fig:fig-1}.  A number of models have been proposed to
explain the structure of the away side peak, including Mach cone/sonic
boom scenarios\cite{Stoecker:2005a,Casalderrey:2006dpa,Ruppert:2005a},
Cherenkov radiation~\cite{Koch:2005sx}, and coupling of jets and
flow~\cite{Armesto:2004vz}, to name a few.

The key to understanding the sources of energy loss, and therefore the
ability to constrain any of these models, is the need to control the
geometry of the collision.  Centrality selection constrains the energy
density of the collision, but that is not quite sufficient.  If we
further choose orientations of the trigger particle with respect to
the reaction plane, this has the effect of varying the effective path
length of the jet. The combination of centrality and trigger particle
orientation will constrain the path length dependence of the energy
loss, enabling the ability to untangle jet and medium responses.

\section{PHENIX Measurements}\label{techno}  

Data taken from three separate years of RHIC operation have been
studied as a function of trigger particle orientation: \AuAu at
$\snn=200~\GeVc$ collisions from Run-4 and Run-7 (taken during 2004
and 2007 respectively), and \CuCu at $\snn=200~\GeVc$ collisions from
Run-5 (taken during 2005).

The PHENIX detector~\cite{Morrison:1998qu} consists of mid-rapidity
spectrometers covering $\left|\eta\right|<0.35$ and forward arms with
acceptance of $1.2<\left|\eta\right|<2.4$.  Hadrons are measured in
the central arms using drift chambers, a ring-imaging cerenkov, and
electromagnetic calorimetry.  For heavy-ion collisions, centrality is
determined from the correlation between the beam-beam counters (BBCs,
located at $3.0<\left|\eta\right|<4.0$) and zero-degree calorimeters
(\AuAu, Run 4) or from the BBCs alone (\CuCu, Run5 and \AuAu, Run 7).
The reaction plane in PHENIX can be measured in multiple complementary
detectors.  For the data presented in this paper, the reaction plane
orientation $\Psi_{RP}$ is determined event-by-event based on the
angular distribution of hits in the BBCs (Run 4 and Run 5) or the
Reaction Plane Detector (installed prior to Run 7), using the standard
method described elsewhere\cite{Adler:2006bw}.  Non-flow effects on
the reaction plane determination have been studied, and it has been
estimated that little to no bias exists in the current
measurements~\cite{:2008cq}.

To determine the jet-induced angular distributions, PHENIX adopts a
two-source model for the source of correlations: hard scattering
(jets) and collective behavior (elliptic flow, characterized by
\vTwo).  Extraction of the jet function in these collisions requires
careful subtraction of the background.  The technique used in the data
presented is the Zero Yield at Maximum (ZYAM)\cite{Adler:2005ee}.  The
key feature of ZYAM is that it assumes there is a value of \dphi where
the jet yield is zero, and uses that point to constrain the background
(\vTwo modulation) level.  To study the dependence on orientation with
respect to the reaction plane, the trigger particles have been placed
into six and three $\dphirp=\phi_{trig}-\Psi_{RP}$ bins from 0 to
$\pi/2$ for \AuAu and \CuCu collisions, respectively.

\section{Results and Discussion}\label{results}

\begin{figure}
\begin{center}
\includegraphics[width=0.95\textwidth]{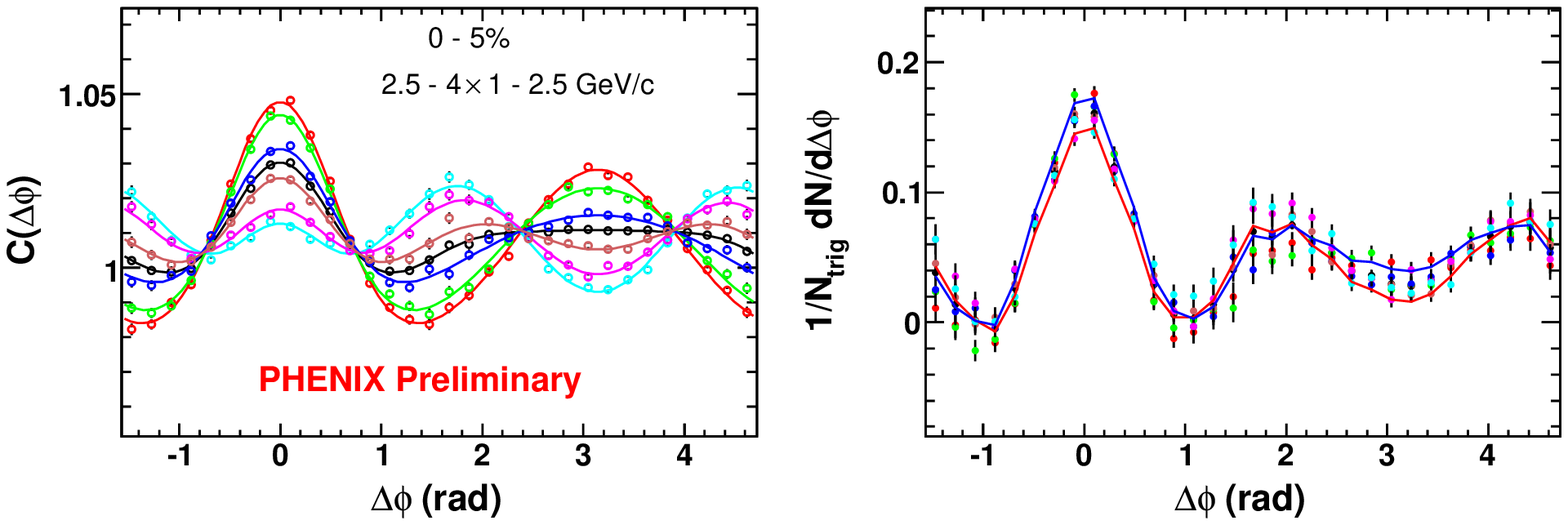}
\includegraphics[width=0.95\textwidth]{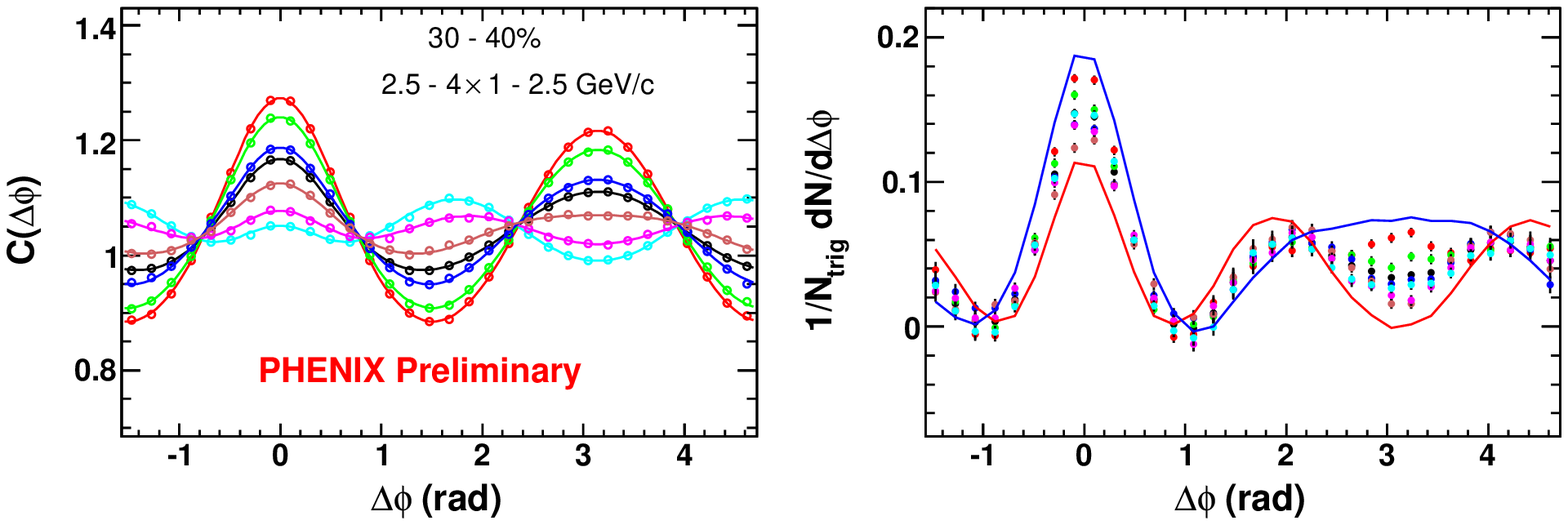}
\end{center}
\vspace*{-14pt}
\caption{Correlation functions (left) and jet-induced yields (right)
  for 0-5\% and 30-40\% centralities in \AuAu collisions at
  $\snn=200~\GeVc$.  Data are from Run-4.}
\label{fig:run4_results}
\vspace*{-14pt}
\end{figure}

\begin{figure}
\begin{center}
\includegraphics[width=0.95\textwidth]{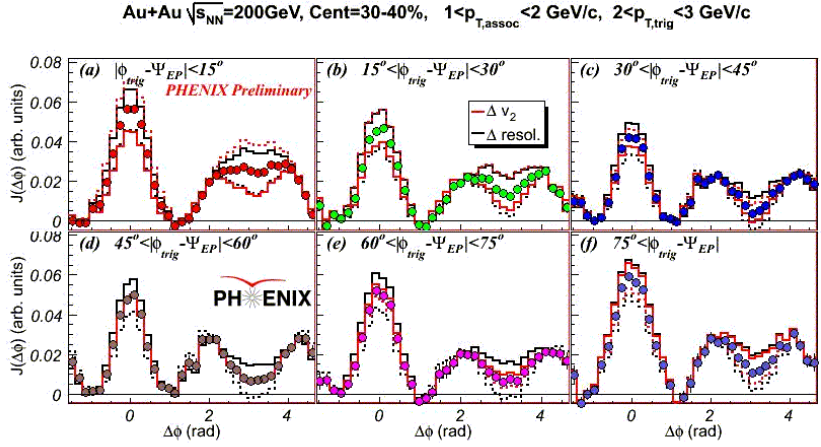}
\end{center}
\vspace*{-14pt}
\caption{Jet-induced yields for 30-40\% centrality in \AuAu collisions
  at $\snn=200~\GeVc$. Data are from Run-7.}
\vspace*{-10pt}
\label{fig:run7_results}
\end{figure}

\begin{figure}
\begin{center}
\includegraphics[width=0.45\textwidth]{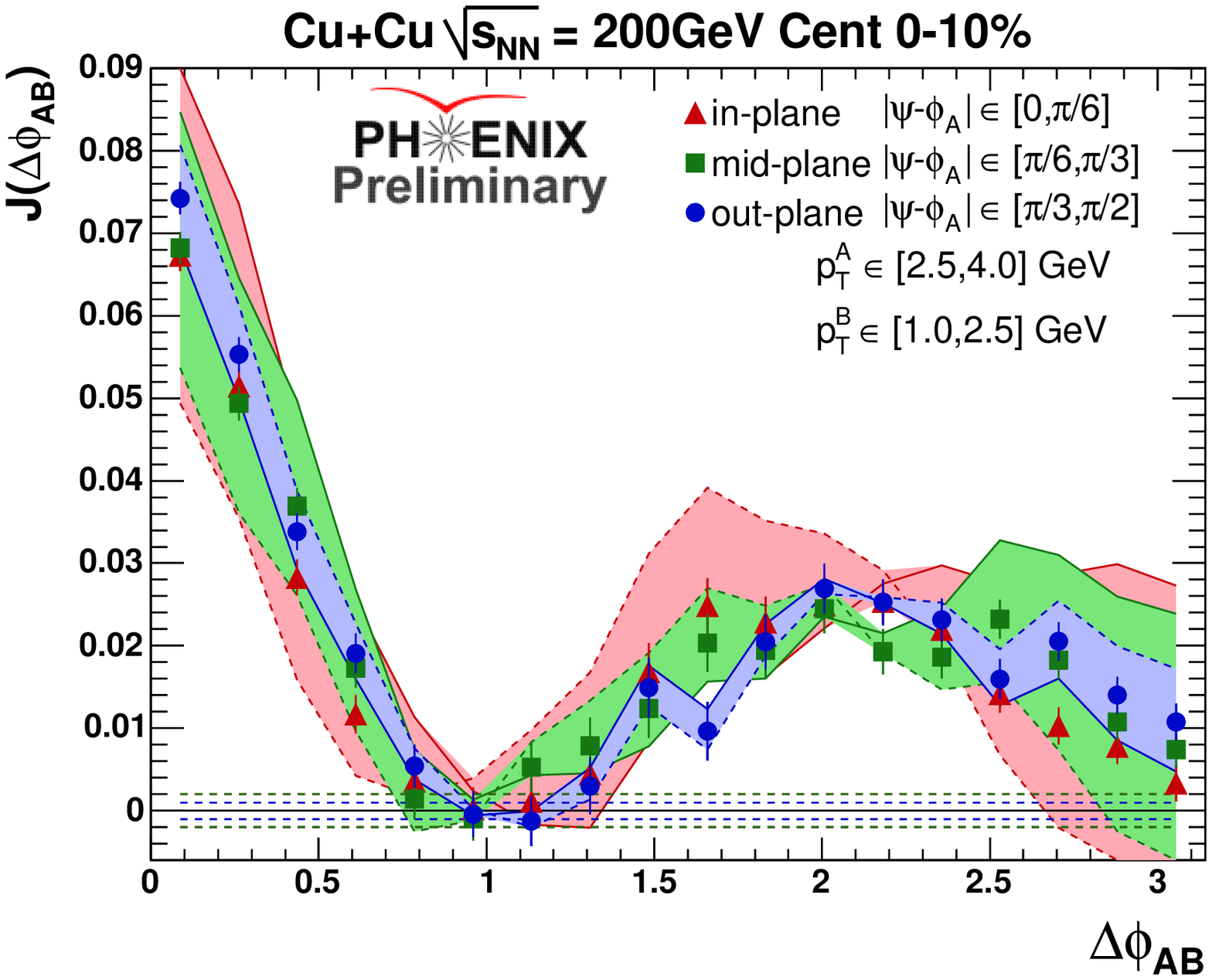}
\includegraphics[width=0.45\textwidth]{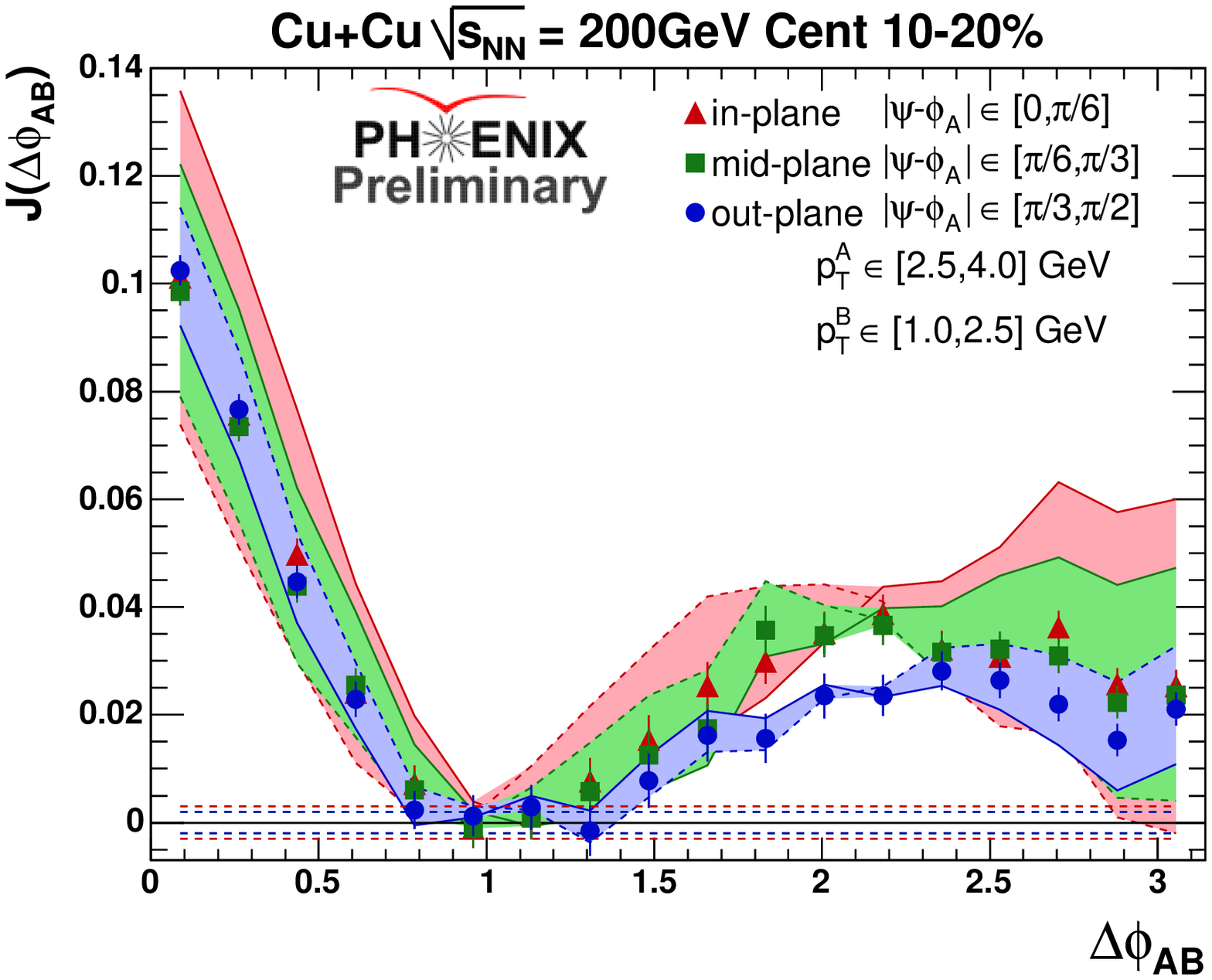}
\end{center}
\vspace*{-14pt}
\caption{Jet-induced yields in \CuCu collisions at
$\snn=200~\GeVc$. Data are from Run-5.}
\vspace*{-10pt}
\label{fig:cucu_jet_rp}
\end{figure}

Run-4 \AuAu correlation and jet functions measured for 0-5\% and
30-40\% events are shown in Figure~\ref{fig:run4_results}.  For the
correlation functions the different curves correspond to the trigger
particle orientation, from 0 (red, or the curve corresponding to the
maximum value at $\dphi=0$) to $\pi/2$ (cyan, or the curve with
minimum value at $\dphi=0$). Several conclusions can be drawn
immediately.  First, there is significant dependence on trigger
orientation in the correlation functions, but after subtraction, there
is little evidence of any dependence.  Also, the same trends in the
data can be seen in both central and semi-central events.  In
particular the shoulder that emerges after subtraction remains the
same, independent of either centrality or trigger orientation.  These
features persist in the Run-7 results, which benefit from higher
statistics and higher resolution reaction plane measurement, as seen
in Figure~\ref{fig:run7_results}.  These trends have been reported for
\AuAu collisions observed at STAR\cite{Feng:QM2008} as well.  Similar
observations have been made using \CuCu jet functions, shown in
Figure~\ref{fig:cucu_jet_rp}.  Like the \AuAu data, there is no
observed dependence of the away-side shape.  Within errors, the
location of the excess is independent of centrality and trigger
orientation.

\begin{figure}
\begin{center}
\includegraphics[width=0.3\textwidth]{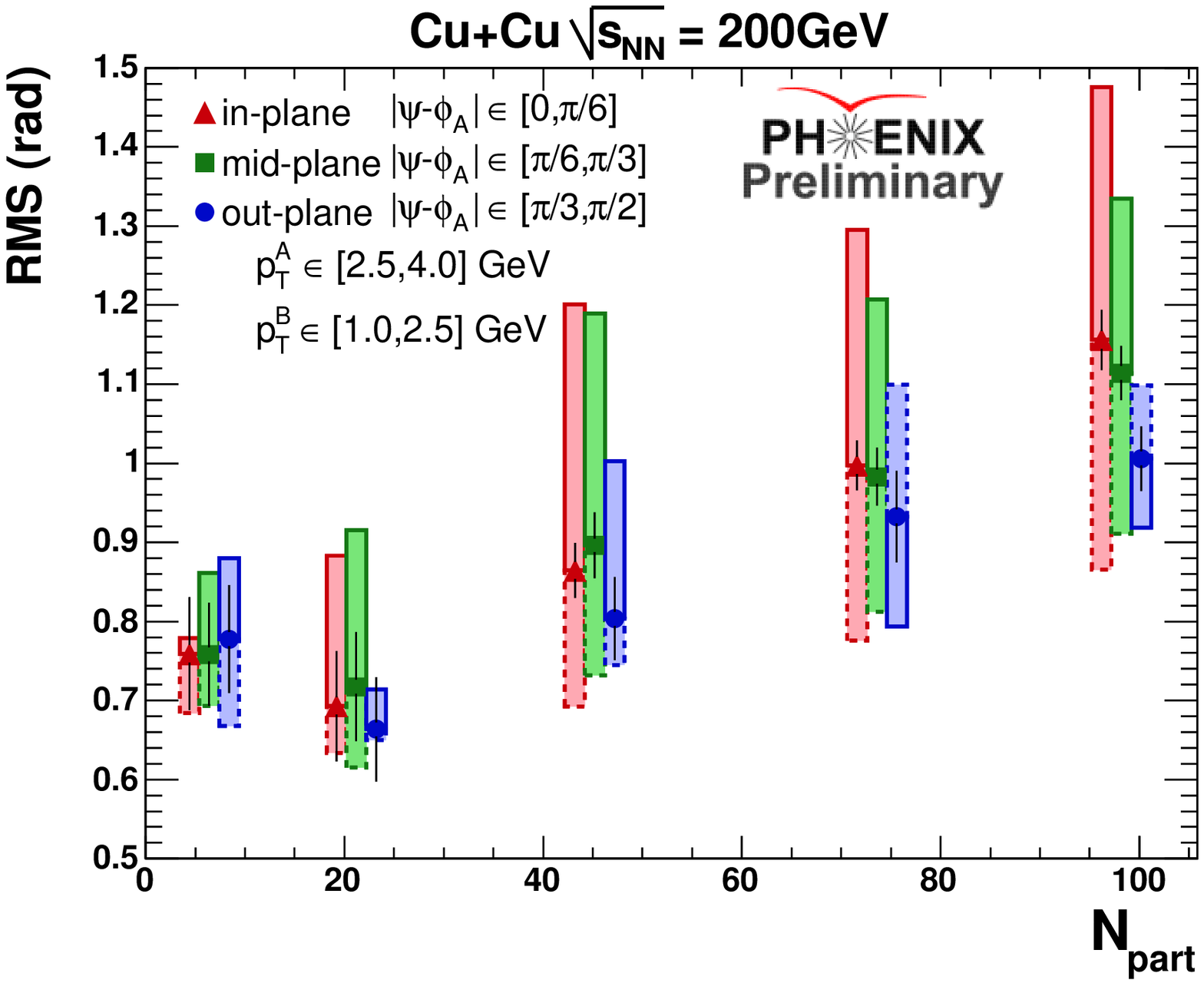}
\includegraphics[width=0.3\textwidth]{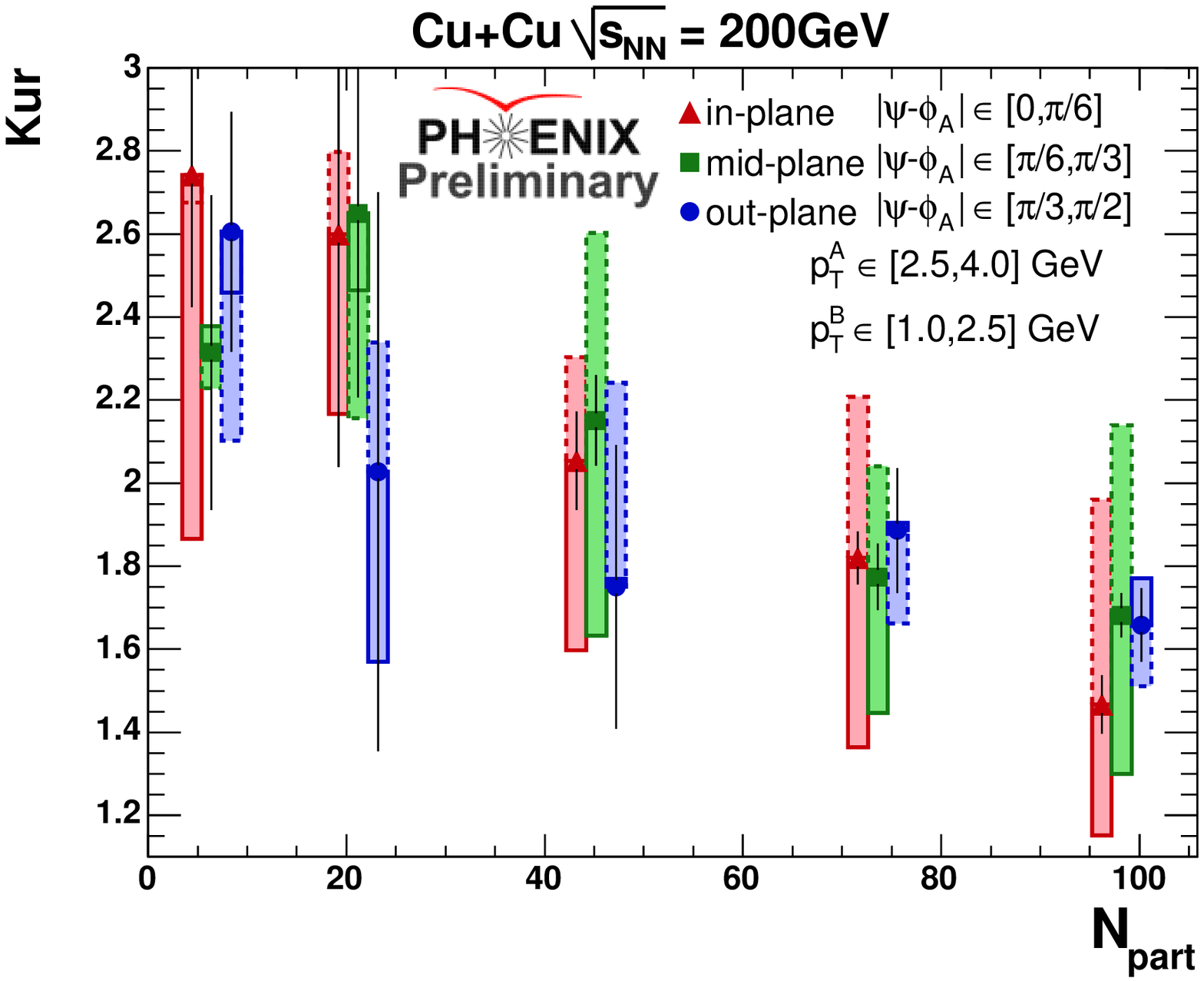}
\includegraphics[width=0.3\textwidth]{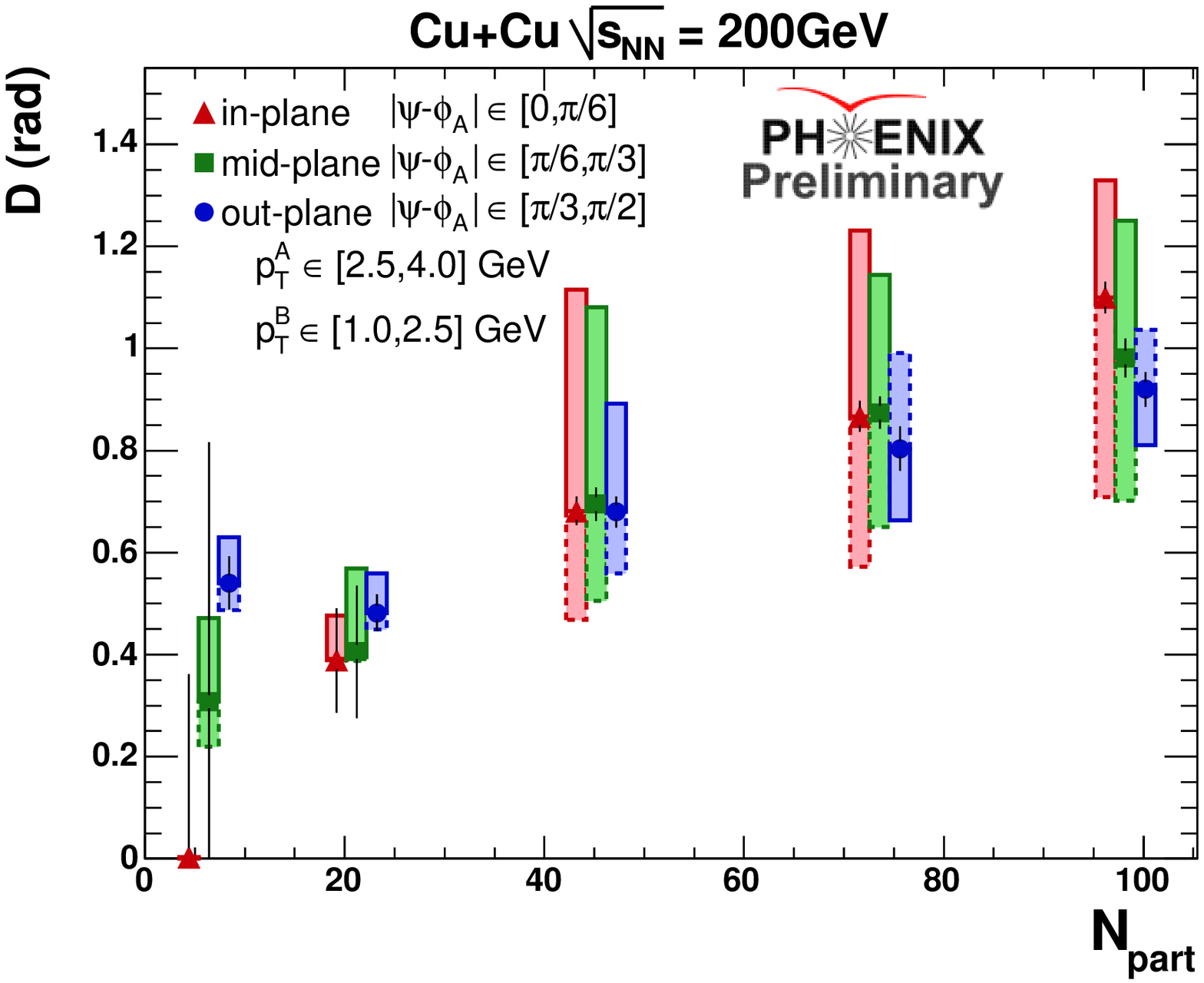}
\end{center}
\vspace*{-14pt}
\caption{(Left to right) The away-side RMS, kurtosis, and $D$ shape
  parameters for \CuCu collisions as a function of centrality, in
  different \dphirp bins.}
\vspace*{-14pt}
\label{fig:cucu_shape_pars}
\end{figure}

The persistance of the away-side shape and the shoulder location in
\dphi has important consequences on energy loss models. The shoulder
region is insensitive to the path length variation, thus disfavoring
jet-flow coupling scenarios.  This feature, however, does tend to
support mach-cone models.  The head region ($\dphi \sim 2\pi$) does
appear to be sensitive to the path length, which is consistent with
energy loss expections. Additionally, it has been shown that the
away-side shape parameters (RMS, kurtosis, and $D$) are independent of
system size and saturate for $N_{coll}>100$\cite{Adare:2006nr}.  This
observation both disfavors Cherenkov gluon radiation and offers
constraints to other models, such as large-angle non-Cherenkov
radiation.  Figure~\ref{fig:cucu_shape_pars} shows the dependence on
trigger orientation for the aways-side RMS, kurtosis, and $D$
parameters for \CuCu collisions at $\snn=200~\GeVc$.  Within
systematics, there is no reaction-plane dependence.

\begin{figure}
\begin{center}
\includegraphics[height=4.1cm]{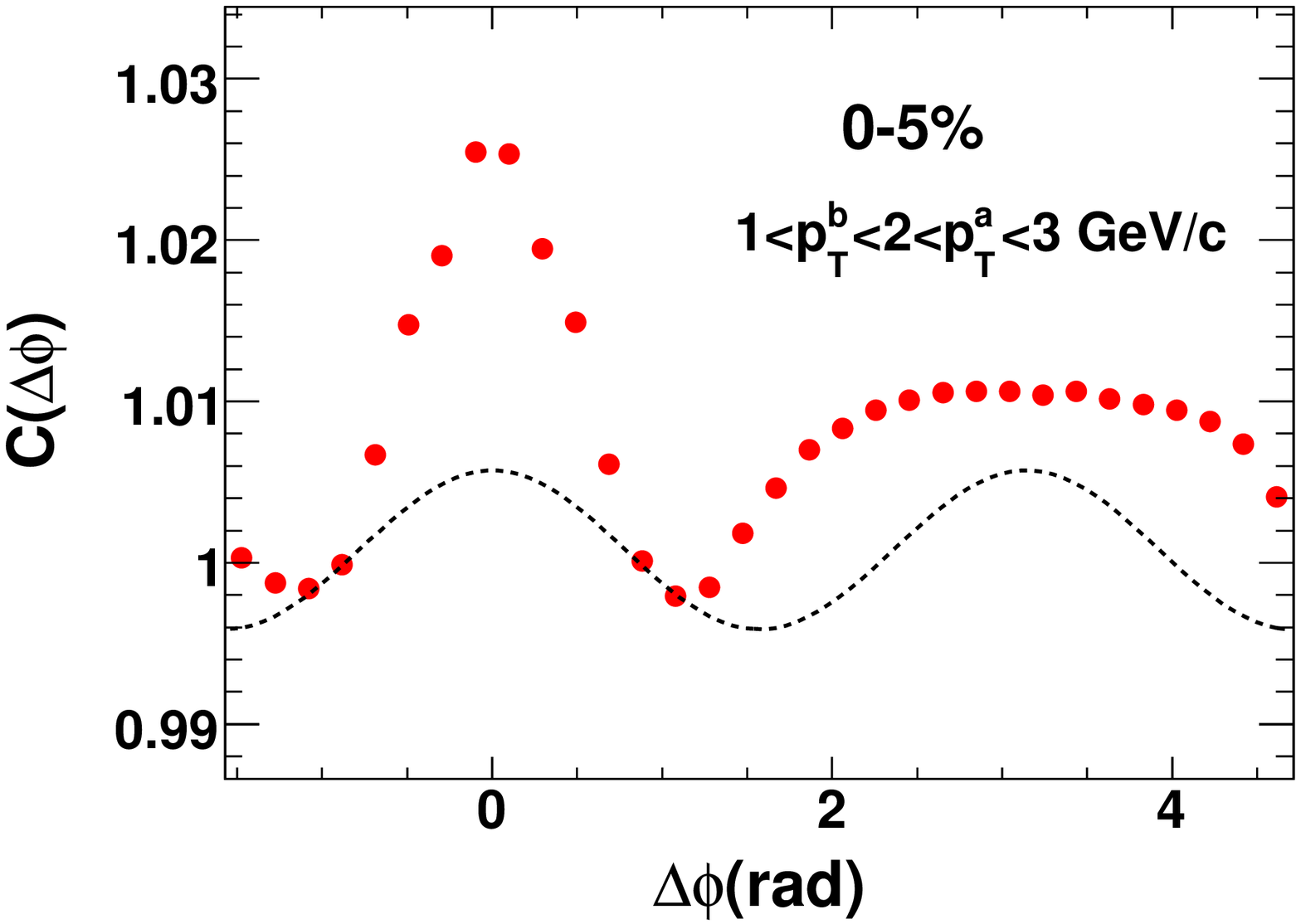}
\includegraphics[height=3.8cm]{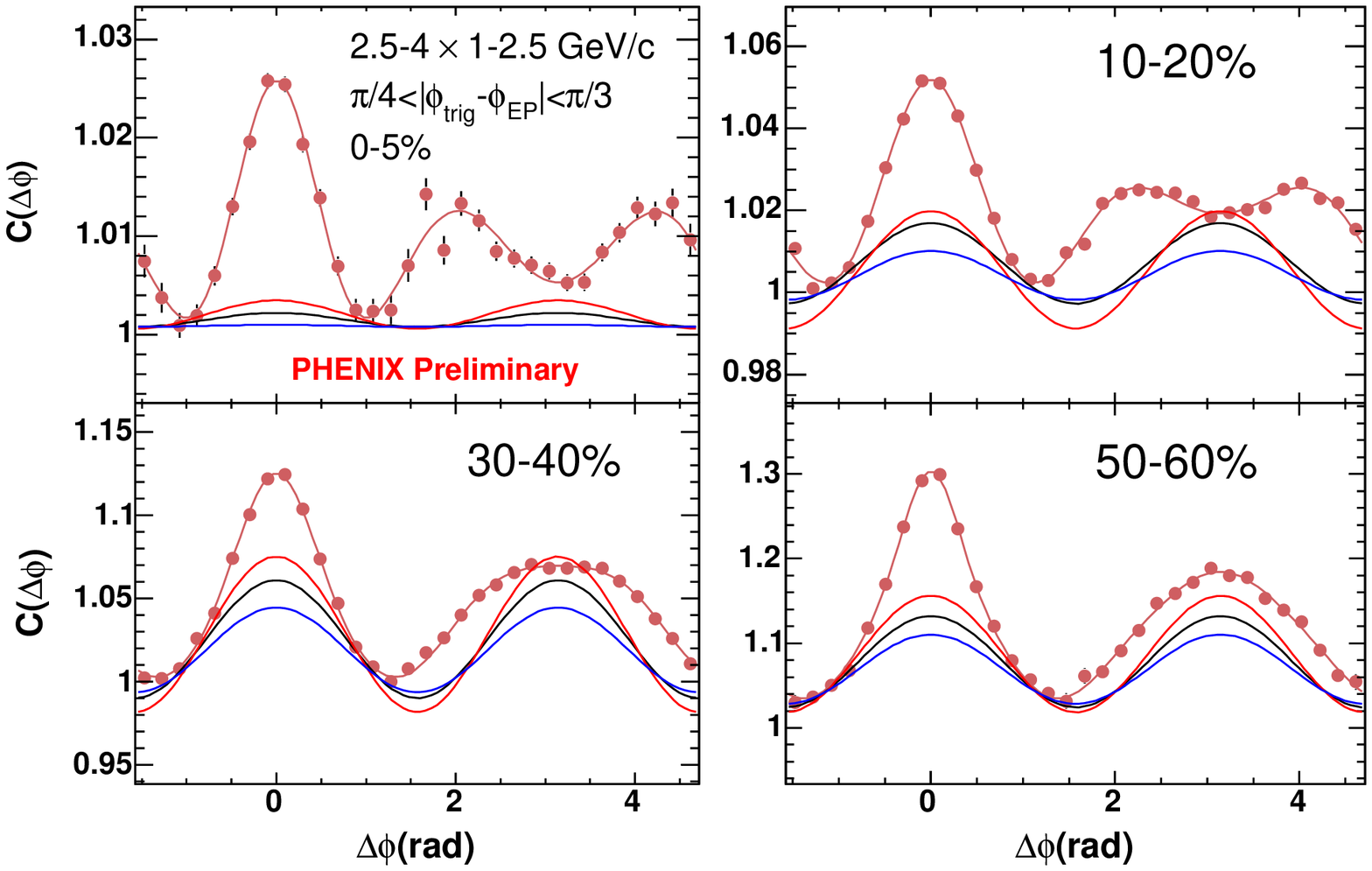}
\end{center}
\vspace*{-14pt}
\caption{Left: Inclusive correlation function for 0-5\% \AuAu
  collisions.  Right: correlation functions as a function of
  centrality for trigger orientation $\pi/3<\dphirp<\pi/4$. The curves
  show the strength of the \vTwo along with the bounds of uncertainty
  on the \vTwo determination.}
\vspace*{-14pt}
\label{fig:dip_example}
\end{figure}

The ZYAM method admittedly has its limits; there is no a priori reason
to the believe that the assumption of zero yield is a particularly
accurate one.  However, even with this caveat, it is impossible to
ignore the non-trivial shape of the away-side yields.  This is because
an excess in the shoulder region is observed even before subtraction,
as seen in Figure~\ref{fig:dip_example}.  To make this point as clear
as possible, in the left panel we have chosen a trigger orientation
that should suffer from the least bias $\vTwo$ determination.  Shown
are correlations in several centrality bins.  Clearly even before
subtraction, a significant excess is seen at $\dphi \sim 2$.

\section{Conclusions}\label{concl}

In summary, the heavy-ion data measured at RHIC have displayed a
evolution of the away-side peak shape with \pT.  A key to
understanding these trends is using the orientation of the trigger
particle in azimuthal correlations to constrain the effect path length
of the fast partons traversing the medium.  Recent measurements at
PHENIX show that the shoulder of the away-side peak, observed at low
to intermediate \pT, appears at the same position ($\dphi \sim 2$),
regardless of the centrality, system size, and orientation of the
trigger particle with respect to the reaction plane.  These data then
provide critical contraints and insights into understanding the
various energy loss models proposed as explanations for the trends
seen in the away-side measurements.
 
 
 
 
\bibliographystyle{spadre2008}
\bibliography{WWND2008_Winter_Proceedings}

\begin{thebibliography}{10}
\expandafter\ifx\csname url\endcsname\relax
  \def\url#1{{\tt #1}}\fi
\expandafter\ifx\csname urlprefix\endcsname\relax\def\urlprefix{URL }\fi

\bibitem{:2008cq}
A.~Adare et~al.  (2008). Submitted for publication in {\it Phys. Rev.} C.

\bibitem{Adler:2006sc}
S.~S. Adler et~al., {\it Phys. Rev.\/} {\bf D74} (2006) 072002.

\bibitem{Adare:2006nr}
A.~Adare et~al., {\it Phys. Rev. Lett.\/} {\bf 98} (2007) 232302.

\bibitem{Adler:2005ee}
S.~S. Adler et~al., {\it Phys. Rev. Lett.\/} {\bf 97} (2006) 052301.

\bibitem{Stoecker:2005a}
H.~Stoecker, {\it Nucl. Phys.\/} {\bf A750} (2005) 121.

\bibitem{Casalderrey:2006dpa}
J.~Casalderrey-Solanaa, E.~Shuryak and D.~Teaney, {\it Nucl. Phys.\/} {\bf
  A774} (2006) 577.

\bibitem{Ruppert:2005a}
J.~Ruppert and B.~Muller, {\it Phys. Lett.\/} {\bf B618} (2005) 123.

\bibitem{Koch:2005sx}
V.~Koch, A.~Majumder and X.-N. Wang, {\it Phys. Rev. Lett.\/} {\bf 96} (2006)
  172302.

\bibitem{Armesto:2004vz}
N.~Armesto, C.~A. Salgado and U.~A. Wiedemann, {\it Phys. Rev.\/} {\bf C72}
  (2005) 064910.

\bibitem{Morrison:1998qu}
D.~P. Morrison et~al., {\it Nucl. Phys.\/} {\bf A638} (1998) 565.

\bibitem{Adler:2006bw}
S.~S. Adler et~al., {\it Phys. Rev.\/} {\bf C76} (2007) 034904.

\bibitem{Feng:QM2008}
A.~Feng, Away-side modification and near-side ridge relative to reaction plane.
  Talk presented at {QM}2008.

\end{thebibliography}

\vfill\eject
\end{document}